\documentclass[useAMS,usenatbib]{mn2e}
\usepackage{graphicx}
\usepackage{subfigure}
\usepackage{bm}
\usepackage{amsmath}
\usepackage{fixltx2e}
\usepackage{xcolor}
\usepackage{amssymb}

\makeatletter

\newcommand{\Rmnum}[1]{\expandafter\@slowromancap\romannumeral #1@}
\makeatother

\title[Sound-Triggered Collapse of Stably Oscillating Low-Mass Cores in a Two-Phase Interstellar Medium]{Sound-Triggered Collapse of Stably Oscillating Low-Mass Cores in a Two-Phase Interstellar Medium}
\author[Ui-Han Zhang, Hsi-Yu Schive and Tzihong Chiueh]{Ui-Han Zhang$^{1}$,Hsi-Yu Schive$^{1}$ and Tzihong Chiueh$^{1}$$^{,}$$^{2}$$^{,}$$^{3}$\thanks{E-mail:
chiuehth@phys.ntu.edu.tw}\\
$^{1}$Department of Physics, National Taiwan University, 10617, Taipei, Taiwan\\
$^{2}$Institute of Astrophysics, National Taiwan University. 10617, Taipei, Taiwan\\
$^{3}$Center for Theoretical Sciences, National Taiwan University, 10617, Taipei, Taiwan}
\begin{document}

\date{Accepted 2015 March 4. Received 2015 February 16; in original form 2014 July 21}

\pagerange{\pageref{firstpage}--\pageref{lastpage}} \pubyear{2015}

\maketitle

\label{firstpage}

\begin{abstract}
\label{abstract}
Inspired by Barnard 68, a Bok globule, that undergoes stable oscillations, we perform multi-phase hydrodynamic simulations to analyze the stability of Bok globules. We show that a high-density soft molecular core, with an adiabatic index $\gamma=0.7$ embedded in a warm isothermal diffuse gas, must have a small density gradient to retain the stability. Despite being stable, the molecular core can still collapse spontaneously as it will relax to develop a sufficiently large density gradient after tens of oscillations, or a few $10^7$ years. However, during its relaxation, the core may abruptly collapse triggered by the impingement of small-amplitude, long-wavelength ($\sim6-36$ pc) sound waves in the warm gas. This triggered collapse mechanism is similar to a sonoluminescence phenomenon, where underwater ultrasounds can drive air bubble coalescence. The collapse configuration is found to be different from both inside-out and outside-in models of low-mass star formation; nonetheless the mass flux is close to the prediction of the inside-out model. The condition and the efficiency for this core collapse mechanism are identified. Generally speaking, a broad-band resonance condition must be met, where the core oscillation frequency and the wave frequency should match each other within a factor of several. A consequence of our findings predicts the possibility of propagating low-mass star formation, for which collapse of cores, within a mass range short of one order of magnitude, takes place sequentially tracing the wave front across a region of few tens of pc over $10^7$ years.
\end{abstract}


\section{Introduction}
\label{sec:introduction}

Interactions of molecular clouds in the interstellar medium (ISM) with shock waves \citep*{bSt, bKl, bXu, bNa2} and turbulence \citep*{bVa, bMa, bLi, bNa1} have been extensively studied in the past decades. The general consensus from these studies is that clouds are most likely to be destroyed by non-radiative shocks due to Kelvin-Helmholtz and Rayleigh-Taylor instabilities, but can be severely compressed leading to gravitational collapse by weak radiative shocks; moreover turbulence generated by proto-stellar winds is capable of self-regulation for star formation. While these results are valid in highly active regions of ISM, such as in giant molecular clouds where nearby proto-stellar winds, supernova explosion and ionization bubbles are at work, the majority part of ISM may on the other hand experience only mild disturbances originated from distant active regions. Here small starless cores of few solar masses in relatively quiescent environments are likely much more abundant in the ISM to contribute to the low-mass end of the stellar initial mass function. In particular, B68, a Bok globule, has been identified to undergo oscillations \citep*{bLada, bRe}, and it is for this population of perturbed low-mass cores that the present work aims to focus. 

Theoretical studies of low-mass star formation in quiescent environments have historically been analyzed with two different model configurations --- the outside-in collapse \citep{bLa, bPe, bHu} and the inside-out collapse \citep{bSh1}. The outside-in collapse model considers an initial uniform cloud that just exceeds the Jeans mass and the collapse is initiated near the cloud boundary where the collapse front propagates from outside toward the center. The inside-out collapse model on the contrary considers an initial cloud already relaxed to a singular isothermal sphere configuration and the collapse is triggered at the center where the collapse front subsequently propagates outward much like an avalanche. Weak disturbances in ISM were never considered to be important to the primary physical mechanisms those works put forth. In this work, we shall instead demonstrate that a small-amplitude, long-wavelength sound may trigger the collapse of an otherwise stable core. This new mechanism involves two key components: a soft equation of state in the core and the sonoluminescence-type resonant sound absorption.

It is well known that in a molecular cloud containing dust and metal, cooling is dominated primarily by molecular line emissions and heating provided by cosmic ray bombardments. While the latter depends linearly on the hydrogen density, the former depends more strongly and complicatedly on the hydrogen density. The different density dependence of cooling and heating yields a peculiar equation of state, with a drop in the adiabatic index $\gamma$ below $1$ for a hydrogen number density exceeding several times $10^3 /cm^3$ \citep{bSp}. Observation of starless clumps also supports a decreasing temperature toward the high-density core \citep*{bBa}. Therefore the molecular core must have a rather soft equation of state. 

On the other hand, sonoluminescence originally refers to underwater air bubbles that undergo radical compression to emit optical lights when subject to the impingement of sounds \citep*{bBr}. When a planar sound wave in water encounters a soft air bubble, it can become a spherical wave propagating into the bubble with an increasing amplitude. Mismatch in sound speeds may however produce wave reflection, but when the sound is in resonance with the air bubble, the wave can be absorbed with high efficiency. It is in this general context that we report this mechanism to be possibly operative in the ISM molecular cores.

We describe our simulation details in Sec. (\ref{sec:numerical scheme and initial condition}). The stability condition of a soft molecular core is derived in Sec. (\ref{sec:stability of a soft core}). Sec. (\ref{sec:core collapse triggered by small-amplitude background sound waves}) presents our main results, addressing the interaction of a small-amplitude wave and a single core. Simulation on the propagating star formation is demonstrated in Sec. (\ref{sec:multi-core collapse triggered by a single wave}), where a single propagating wave encounters several cores. We give conclusions in Sec. (\ref{sec:conclusions}).

\section{Numerical Scheme, Equation of State and Initial Condition}
\label{sec:numerical scheme and initial condition}

We demonstrate the core collapse mechanism numerically. Our simulations are conducted with the GAMER code, a graphic-process-unit (GPU) accelerated hydrodynamics adaptive-mesh-refinement (AMR) code, which outperforms the CPU code by a factor of several tens in computation speed \citep*{bSc1, bSc2}. The GAMER code solves ideal hydrodynamics with self-gravity in conservation forms, and it has installed several solvers, such as CTU (corner transport upwind), MHM (MUSCL-Hancock) and RTVD (Relaxing Total Variation Diminishing). The RTVD solver is chosen for this problem. The scheme adopts an approximate measure for estimating the information speed, making it convenient to handle an unusual equation of state adopted in this work. With GAMER, we may search for the nonlinear stability boundary with hundreds of 3D simulations, much like Monte-Carlo simulations, in a sizable simulation box. We are also able to achieve $(3.145728\times10^6)^3$ effective resolution when the core collapses.

A simplified initial bare core is adopted, which is to be relaxed to an appropriate configuration. This initial dense H\textsubscript{2} core has 0.2 pc diameter, $\sim 20$ K temperature and $\sim 10^4$ cm$^{-3}$ atomic hydrogen density, surrounded by a voluminous H\Rmnum{2} gas of $\sim 10^4$ K temperature and $\sim10$ cm$^{-3}$ density and separated by an infinitely thin transition layer.

Our simulation domain is a 37 pc cube with 0.048 pc base grid resolution. The base grid resolution is defined as level 0 and the $n$-th refinement is level $n$, with level $n$ twice higher resolution than level $n-1$. The space position and the density are used as the refinement criteria. For the space position, any grid with distance from the core center smaller than $2^{5-n}R_c$ is refined to level $n$, where $n=1$ - $4$ and $R_c$ is the initial core radius. This condition guarantees the core is embedded in grids of sufficiently high resolution. On the other hand, the grid with density higher than $37.5*4^{n-1}\rho_g$ is also refined to level $n$, where $n =1$ - $12$ and $\rho_g$ is H\Rmnum{2} mass density. This condition aims to capture the collapse center. When the collapse center reaches a density that yields a Jeans length equal to twice the finest grid size, we terminate the simulation; typically the density dynamical range is $10^{10}$. Periodic condition is employed throughout this study.

We treat the two-phase gas as a single-component gas with an unusual equation of state. The index of adiabaticity $\gamma$ approaches $0.7$ in the interior of the molecular core and $1.1$ for the almost isothermal warm H\Rmnum{2} gas. For any gas dynamics simulation code written in a conservation form, the internal energy density $e$ and the pressure $P$ are both needed as the dynamical variables. We model the equation of state as $P=a(s)g(\rho)$, and the squared sound speed $C_s^{2}=\gamma(\rho) (P/\rho)$, where $\rho$ the mass density, $a(s)$ the entropy function, $s$ is the entropy and the adiabatic index $\gamma(\rho)\equiv {{d \ln g(\rho)}/d\ln \rho}$. To find the internal energy density, $e$, from the pressure $P$, the first law of thermodynamics demands,
\begin{equation}
\label{equ:the first law of thermodynamics 1}
{de\over\rho}-{{e+P}\over{\rho^{2}}}d\rho=0,
\end{equation}
and the internal energy density $e$ is thus related to the pressure $P$ through
\begin{equation}
\label{equ:internal energy density as the function of pressure and density}
e=P\left[{\rho\over{g(\rho)}}\int_{0}^{\rho}{g(x)\over x^{2}}dx\right].
\end{equation}

To model the change in the adiabatic index in the two-phase gas, we first let $g(\rho)=({\rho/{\rho_{0}}})^{\bar\gamma(\rho)}$, and given the limiting values of $\bar\gamma(\rho)$ in the core and in the warm gas, we model
\begin{equation}
\label{equ:adiabatic index as function of density}
{\bar\gamma}({\rho})={\bar\gamma}_{0}-{\Delta}{\bar\gamma}\tanh\left({{\ln\left({{\rho}\over{\rho_{0}}}\right)}\over{\eta}}\right),
\end{equation}
where the core density ${\rho}_{c}={\rho_{0}}\exp(1.02\eta)$ and the warm gas density ${\rho}_{g}={\rho_{0}}\exp(-1.07\eta)$. Here, $\rho_0$ is the density at the transition boundary of the two-phase gas, and $\eta$ represents the range of transition density in unit of $\rho_0$, chosen to be ${\eta}=3.3$ to yield $\rho_{c}/\rho_{g}\approx 1000$. Finally, we set $\bar\gamma_{0}=0.9$ and $\Delta\bar\gamma=0.2$.  (Note that $\gamma(\rho)$ is different from $\bar\gamma(\rho)$ due to an extra contribution from $d\bar\gamma/d\rho$.)

\begin{figure}
\includegraphics[scale=0.355, angle=270]{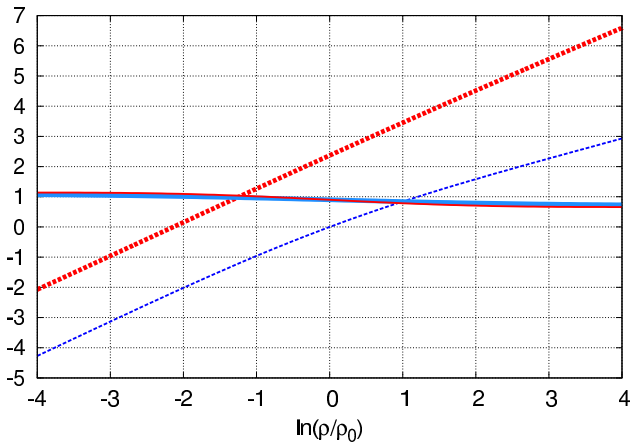}
\caption{The density dependence of $\ln(e/a(s))$ (red dashed line), $\ln(P/a(s))$ (blue dashed line), $\gamma({\rho})$ (red solid line) and ${\bar\gamma}({\rho})$ (blue solid line).}
\label{fig:equation_of_state}
\end{figure}

Fig. (\ref{fig:equation_of_state}) depicts various thermodynamics quantities, $P/a(s), e/a(s), \gamma$ and $\bar\gamma$, as a function of density. Note that $\gamma$ and $\bar\gamma$ closely track each other, and the ratio between $e$ and $P$ increases with the density due to an increasing heat capacity.

We start the simulation with the aforementioned uniform-density, quasi-spherical H\textsubscript{2} core of mass $1.5$ $M_\odot$ and temperature $20$ K, which is in pressure balance with the surrounding warm gas of $10^4$ K. (A "quasi-sphere" here is a sphere superposed with about $20\%$ of higher multi-pole moments.) So long as the core size is smaller than the Jeans length, a uniform core is stable, and the core oscillates with large amplitudes since the gravitational force is initially unbalanced. The oscillation period is about $2$ Myr. 

\begin{figure}
\includegraphics[scale=0.355, angle=270]{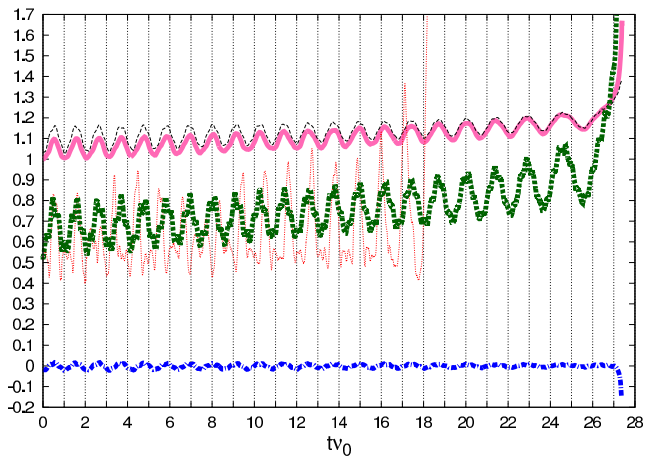}
\caption{Time evolution of the central densities of quasi-spheroid (thick dashed line) and ellipsoid (thin dotted line) normalized to $10^4\rho_g$, the density gradient $A(t)/A(0)$ evaluated by Eq. ({\ref{equ:core mass relation}})(thick solid line), the first two terms (thin dashed line) and all terms (thick dash-dotted line) on the right of Eq. (\ref{equ:virial theorem}). The core oscillates and relaxs to a final configuration accessible to collapse.}
\label{fig:density_peak_and_virial_condition_evolution_with_caption}
\end{figure}

The core can quickly adjust to establish a pressure gradient within the initial 10 oscillation periods to balance the gravity. The central temperature of the core by now drops from $20$ K to $12$ K and the central density rises from $10^4$ cm$^{-3}$ to $5\times10^4$ cm$^{-3}$. After that, the enhancement of the central density slows down while the core continues to oscillate with finite amplitudes. When the central density becomes sufficiently large after about $100$ oscillation periods, the core becomes unstable and spontaneously collapses. The evolution of the central density in Fig. (\ref{fig:density_peak_and_virial_condition_evolution_with_caption}) shows the last tens of oscillations prior to the spontaneous collapse.

The peculiar feature of a strong density gradient for driving the core instability will be discussed in the next section. For now we shall address the mechanism by which changes of the density gradient are possible.

An isentropic equilibrium with spherical symmetry can only support potential flows, and it cannot produce vortex flows to yield material mixing. Hence the density gradient cannot change securely. However, an equilibrium sphere with an existing entropy gradient may generate vortical motion and can lead to material mixing. Rayleigh-Taylor instability is a familiar example. In our case, the system has an entropy gradient at the core boundary. But it is Rayleigh-Taylor stable since the cold matter is at the bottom of the gravitational potential instead of on the top. Nevertheless, we have a core on one hand possessing high-order multipole moments and on the other hand oscillating with large amplitudes. Misalignment of the entropy and density gradients can develop as a result of nonlinear coupling of oscillations of different multi-pole moments. Therefore weak vortices are generated, thereby mixing the core over a long time.

\begin{figure}
\includegraphics[scale=0.42]{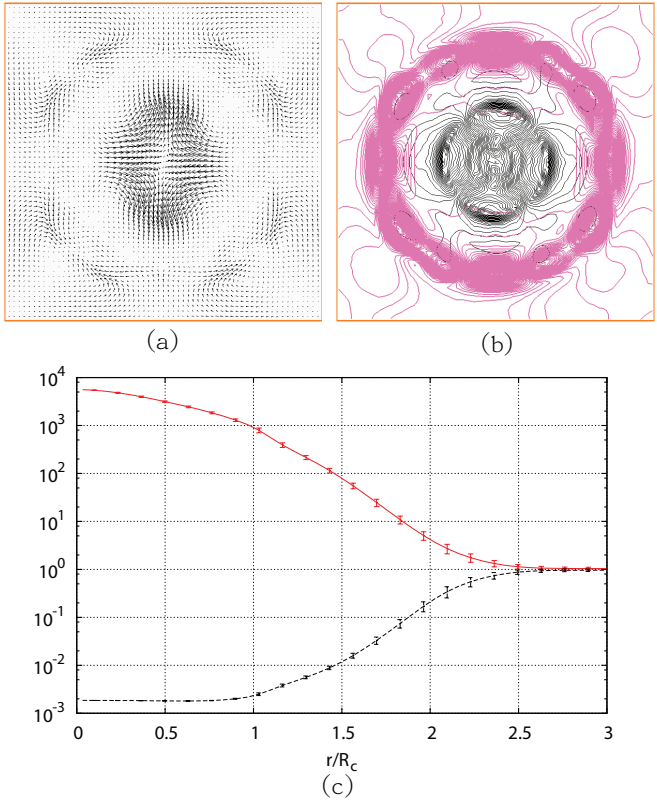}
\caption{(a): A slice of the vortical velocity field normalized to the local sound speed. This slice cuts across the center of the core and has a size of $6R_c$.  The maximum Mach number is about $4\%$. (b): The contour of the high-order multipole moments of density (dark curve) and entropy function $P/g(\rho)$ (light curve) on the same slice of (a). The density and the entropy function are normalized to the background values in the H\Rmnum{2} gas. The density and entropy gradients are the source of oscillating vortex flow and are co-spatial with the vortex flow. (c): The shell-averaged radial profiles of density (solid line) and the entropy function $P/g(\rho)$ (dashed line). The normalizations are the same as (b). The error bars indicate the magnitudes of high-order multipole moments.}
\label{fig:initial_condition}
\end{figure}

To demonstrate, we show a slice of the core through the core center in Fig. (\ref{fig:initial_condition}.a), and depict the two-dimensional vortical velocity normalized to the local sound speed at the moment of 72 oscillation periods. The vortical velocity field is computed by a projection operator ${\bf I}-\hat k\hat k$ acting on the Fourier component of the velocity field, where ${\bf I}$ and ${\hat k}$ are the identity tensor and the directional unit vector of wavenumber ${\bf k}$, respectively. Sub-sonic weak vortex flows are clearly seen, where the Mach number is about few percent. On this slice, the flow exhibits a quadrupole pattern, for which a radial inflow on one axis is accompanied by a radial outflow on another axis, and the flow has a negligible component in the direction of the third axis. Along the third axis, the flow velocity is also small. Thus, the flow is on a thick torus in three-dimensional space, and it has a dominant quadrapole velocity pattern. The direction of the vortex flow can actually reverse and it oscillates quasi-periodically with frequencies comparable to the density oscillation.
 
To investigate density and entropy, the respective spherical shell averages are subtracted from these two quantities. The contours of the instantaneous high-order multipole moments of density and entropy are plotted in Fig. (\ref{fig:initial_condition}.b) on the same slice as Fig. (\ref{fig:initial_condition}.a). The density and entropy gradients are clear in regions where vortex flows are strong, and the misalignment between these two gradients are the source of oscillating vortex flows.
 
In Fig.(\ref{fig:initial_condition}.c), we plot the spherical shell average profiles of density and entropy. The scatters around the average profiles indicate the magnitudes of high-order multipole moments. The initial sharp density and entropy boundary is now smeared over an extended regions after $72$ oscillation periods due to the material mixing, which yields a more amenable configuration for a molecular core. Therefore we shall take the relaxed configuration of Fig.(\ref{fig:initial_condition}.c) as the initial condition for the study to follow.

\section{Stability of A Soft Core}
\label{sec:stability of a soft core}

Despite the very soft equation of state, the core is found to be nonlinearly stable.  This counter-intuitive core stability can be seen from the virial equation,
\begin{equation}
\label{equ:virial theorem}
\left\langle r \rho{\partial^2\xi_r(r,t)\over\partial t^2}\right\rangle= 3{\langle} P{\rangle}-3P(R)+{\langle} {\Omega}{\rangle},
\end{equation}
where the volume average is defined as $\langle f \rangle\equiv \int_0^R f d^3r/ (4\pi R^3/3)$ integrated up to a {\it fixed} radius $R$, which we call $R$-average, $\xi_r(r,t)$ is the radial displacement, $\Omega$ the potential energy density, and $P(R)$ is the pressure evaluated at radius $R$. For convenience, we choose $R$ to be located immediately outside the core radius. With $\bar\gamma <1$, the pressure within the core must decrease upon contraction. However, another contribution to the $R$-average pressure from the warm gas does the opposite.  It increases the $R$-average pressure when the core contracts since the volume occupied by the warm gases within $R$ increases from zero to a positive value. To demonstrate the stability, we compute from the simulation the time evolution of the average density gradient $A$, the first two terms and all terms combined on the right of Eq. (\ref{equ:virial theorem}) in Fig. (\ref{fig:density_peak_and_virial_condition_evolution_with_caption}). As the core contracts, the average density gradient $A$ increases and the other two curves also increase. All three curves oscillate in phase, indicative of that the restoring force associated with the increase of $R$-average pressure is capable of resisting the enhancement of self-gravity and produces stable nonlinear oscillations.

The above analysis of the first-order virial equation for nonlinear stability is at best {\it ad hoc}, largely relying on the simulation result.  As the stability of such a soft core is new and counter-intuitive, we now analyze this problem in a more rigorous manner.  Despite the core configuration in the simulation is away from the dynamical equilibrium, the underlying linear stability of the core must be obeyed to account for the simulation result.  We thus assume the existence of a quasi-static equilibrium and perform a linear stability analysis about this equilibrium. To simplify the analysis, we model the dynamical equilibrium to consist of a cold spherical core possessing a density gradient and a uniform warm diffuse gas that is separated from the core by a sharp boundary.

In a dynamical equilibrium, the first-order virial is identically zero. Variational principle deals with the second-order virial, and the second-order potential energy $\delta^2 W$ of linear perturbations can be written as,
\begin{equation}
\label{equ:stability analysis_1}
\left.\begin{aligned}
\delta^2W\equiv &\int d^3{\bm{r}} [ \gamma P_0(\nabla\cdot {\bm{\xi}})^2-({\bm{\xi}}\cdot \nabla \rho_0)({\bm{\xi}}\cdot\nabla \phi_0)\\
                & +2({\bm{\xi}}\cdot \nabla P_0)\nabla\cdot{\bm{\xi}}+(\rho_0{\bm{\xi}}\cdot\nabla\delta\phi)] \\
                & -\int_S d{\bm{S}}\cdot {\bm{\xi}}[({\bm{\xi}}\cdot\nabla P_0) +\gamma P_0(\nabla\cdot {\bm{\xi}}) ],
\end{aligned}
\right.
\end{equation}
where the surface integral is evaluated at $R$, just outside the core radius $R_c $. The lowest-energy mode must avoid exciting the positive-energy sound wave in the high temperature region, and hence the displacement is incompressible, $\nabla\cdot{\bm{\xi}}\to 0$. Moreover, as the density is low and the gravity negligible, the warm gas has a vanishingly small equilibrium pressure and density gradients, i.e., $\nabla P_0 \to 0$ and $\nabla \rho_0 \to 0$.  The incompressibility of the warm gas also yields a vanishing small density perturbation, thus a vanishingly small gravitational perturbation, $\delta\phi \to 0$.   Finally, the surface integral is also small due again to incompressibility and to the small pressure gradient just outside the core. The second-order potential energy $\delta^2W$ is thus contributed entirely by the region $r\leq R$. 

Like a pulsating star, the most unstable mode involves only the radial displacement, ${\bm{\xi}}=\xi_r \hat{\bm{r}}$, so that it can tap the gravitational energy. Eq. (\ref{equ:stability analysis_1}) then becomes
\begin{equation}
\label{equ:stability analysis_2}
\left.\begin{aligned}
{\delta^2 W}= & 4\pi \int_0^R dr \Big [ {\gamma P_0\over r^2} \left({d(r^2\xi_r)\over dr}\right)^2-r^2\xi_r^2{d\rho_0\over dr}{d\phi_0\over dr} \\
                            & +2\xi_r{dP_0\over dr}{d(r^2\xi_r)\over dr} -r^2\xi_r^2 4\pi G\rho_0^2 \Big ].
\end{aligned}
\right.
\end{equation}
Note that $d\rho_0/dr$ in the second term contains a Dirac-$\delta$ function at the core boundary. This term is positive and results in the stabilizing effect mentioned earlier, i.e., an increase of average pressure within a fixed $R$ contributed by the warm gas to stabilize the perturbations. 

We next parametrize the equilibrium density as $\rho_0(x) =\rho_0(1) (x+A(1-x))$, where $x\equiv r/R$ and $A>0$, i.e., a linear model from which the force balanced pressure can be calculated.  It yields $P_0(x)=\rho_0(1) T_0(1) +\pi G\rho_0(1)^2 R^2[(1/4)+(5/18)A+(5/36)A^2-(2/3)A^2 x^2+(7/9)A(A-1)x^3-(1/4)(A-1)^2 x^4]$.  Here, $A$ can be regarded as the average density gradient of the core. 

Finally we adopt a trial function $r^2\xi_r = x^3-(3/4)x^4-\beta(x^3-(3/5)x^5)$. Here $\beta$ is an optimization parameter, where $\delta^2 W$ is to be minimized with respect to $\beta$. An optimal choice of the trial function can provide an accurate estimate for the marginal stability condition. This trial displacement gives uniform compression at $x=0$ and becomes incompressible at $x=1$. Demanding the radial displacement and its radial derivative to be continuous across the boundary, we let $r^2\xi_r=1/4-(2\beta/ 5)$ outside the core, which is incompressible as desired.

Substituting the above into Eq. (\ref{equ:stability analysis_2}) and setting the first $\beta$-derivative of $\delta^2 W$ equal to zero, we find that the optimal $\beta=\beta_{op}$. For all practical purposes when $\beta_{op}$ is on the order of unity, $\delta^2 W$ is insensitive to $\beta$. Hence for the time being we set $\beta=0$ to simplify the algebra and the presentation. Full expressions due to a finite $\beta$ will be given in Appendix.

Eq. (\ref{equ:stability analysis_2}) becomes
\begin{equation}
\label{equ:stability analysis_3}
\left.\begin{aligned}
& {\delta^2 W}= {\pi\rho_0(1)\over R_c^3} \Big \lbrace {{6 \gamma T_0(1)}\over{5}} + \\
& {{\pi G R_c^2\rho_0(1)}\over{420}} \Big [ \gamma(111+100A+29A^2) - {{(720+695A+217A^2)}\over{6}} \Big] \Big \rbrace,
\end{aligned}
\right.
\end{equation}
where we have let $R=R_c$.  

On the other hand, the core mass, denoted as $M_c$, is a conserved quantity and related to the core radius $R_c$ by the following relation.
\begin{equation}
\label{equ:core mass relation}
R_c = \left[{{3M_c}\over{\pi \rho_0(1)(3+A)}}\right]^{1\over3}.
\end{equation}
We define $\Phi_0(1)\equiv -GM_c/R_c$.
Substituting the Eq. (\ref{equ:core mass relation}) into Eq. (\ref{equ:stability analysis_3}), the unstable condition, where $\delta^2 W$ is negative, becomes,
\begin{equation}
\label{equ:stability_condition}
\left.\begin{aligned}
{T_0(1) \over{|\Phi_0(1)|}} \leq & {1\over 168(3+A)\gamma} \Big [ \Big ( {217\over6}-29\gamma \Big )A^2+ \\
                                & \Big ( {695\over6}-100\gamma \Big )A  + (120-111\gamma) \Big ].
\end{aligned}
\right.
\end{equation}

Note that core relaxation increases the average density gradient $A$. As the right-hand side is an increasing function of positive $A$ for $\gamma$ less approximately than unity, the core, at a fixed $T_0(1)/|\Phi_0(1)|$, therefore becomes more unstable when the density gradient $A$ increases. The ratio $T_0(1)/{|\Phi_0(1)|}$ is measured to be $1.28$ from the simulation at the moment of spontaneous core collapse, and the marginally stable value of $A$ is thus predicted to be $8.9$ for $\gamma=0.7$ according to Eq. (\ref{equ:stability_condition}). (For a finite $\beta$ the marginally stable $A$ is predicted to be $8.75$.) This is to be compared with the measured $A=7$ from the simulation at core collapse. Given the crude linear model with a sharp boundary for the linear stability analysis, the prediction is in fair consistency with the simulation result.

The evolution of $A$ also shown in Fig. (\ref{fig:density_peak_and_virial_condition_evolution_with_caption}) reveals that the core configuration begins with $A=5.6$ and evolves toward collapse when $A=7$ in $28$ initial oscillation periods. Therefore, the small density gradient explains the stability of soft core and may explain the stability of Barnard 68.

The measured $A$ is found by the following procedures. The core mass $M_c$ is first evaluated by the configuration before core relaxation where the core radius $R_c$ is chosen to be the radius of a uniform core. Since $M_c$ is conserved, the core radius $R_c$ can be determined by calculating the enclose mass at any moment. The density at the boundary $\rho_0(1)$ can be measured once $R_c$ is known. Finally, the density gradient $A$ is determined by using Eq. ({\ref{equ:core mass relation}}).

\section{Core Collapse Triggered by Small-Amplitude Background Sound Waves}
\label{sec:core collapse triggered by small-amplitude background sound waves}

Two different cases are to be investigated. First, a standing wave acts on a quasi-spherical core studied in the last section; this case provides a baseline stability criterion. Second, a Gaussian pulse impinges on an elliptical core; this case addresses a more general situation.

Since the core is oscillating, it is natural to examine the interaction efficiency as a function of wave frequency. Here the core is located at the pressure node of the standing wave. When wave frequency coincides with the core frequency, the interaction is expected to be most efficient, thus requiring the lowest wave amplitude to drive the collapse. We define the threshold wave amplitude as the one that can marginally compress the core to collapse.  The marginally driven case is one that the core peak density rises monotonically, but hesitantly, toward collapse in one to two oscillation period.

\begin{figure}
\includegraphics[scale=0.34, angle=270]{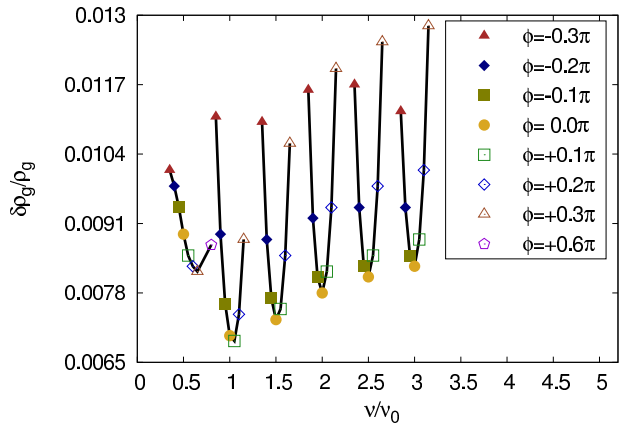}
\caption{Frequency/phase response of the threshold wave amplitude (${\delta}{\rho_g}/\rho_g$) for triggering collapse of a quasi-spheroid core by a sinusoidal standing wave. The six groups of symbols correspond to $\nu/\nu_0=0.5, 1, 1.5, 2, 2.5, 3$, where $\nu$ and $\nu_0$ are the wave and core frequencies, respectively. Different symbols in each group represent different initial relative phases between the two oscillations.}
\label{fig:phase_diagram_standing_wave}
\end{figure}

We find that the resonant interaction is indeed the most efficient; however the quality factor of resonance is rather low, as shown in Fig. (\ref{fig:phase_diagram_standing_wave}). Here, we plot the threshold wave amplitude as a function of the wave frequency and the relative phase between core and wave oscillations. Since collapse occurs in one core oscillation period, one therefore expects the relative phase to be also a crucial parameter. Fig. (\ref{fig:phase_diagram_standing_wave}) contains 6 groups of data points. Data in every group have the same wave frequency, and six frequencies $\nu=0.5, 1, 1.5, 2, 2.5$ and $3$ times of the initial core oscillation frequency $\nu_0$ of Fig. (\ref{fig:density_peak_and_virial_condition_evolution_with_caption}) are investigated. Data in the same group represent different relative phases, and data of the same symbol in different groups represent the same relative phase but different wave frequencies.  

Since wave and core frequencies are different, the relative phase needs to be defined carefully. Here, it is the wave phase when the core oscillation phase is most rarified, or vice versa. The zero relative phase $\phi=0$ corresponds to the situation that the most rarified phases of the two oscillations coincide.  A non-zero $\phi$ refers to the phase of the wave oscillation or the core oscillation whichever is faster.

It is seen that $\phi=0$ is almost always the most efficient phase for all frequencies to drive core collapse. When $\phi=0$ and $\nu=1.0\nu_0$, it requires the lowest wave amplitude to drive core collapse. When $\phi=0$ but $\nu\neq \nu_0$, the mismatch in oscillation timing requires a more vigorous sound amplitude to drive core collapse. However, it turns out that not much higher amplitude is needed to drive the off-resonance collapse, a clear indication of broadband resonance. The 3dB bandwidth of the threshold wave power is about a factor $6$, from $\nu/\nu_0 =0.5$ to $3$, with an allowable range of phase mismatch about $\pm 60 $ degrees. 

We next consider an ellipsoidal core impinged by a Gaussian sound pulse to test whether the above result holds in a more general case. The ellipsoid is parameterized by the lengths of the three principle axes, which are respectively $1.3R_c$, $R_c$ and $R_c/1.3$, and obtained by uniform volume-preseving distortion of the above quasi-spherical core, so that the core mass is preserved. The ellipsoidal core is also stable and undergoes similar relaxation oscillations as the quasi-spherical core, albeit the stable oscillation ends earlier and the amplitude is twice greater.  Fig. (\ref{fig:density_peak_and_virial_condition_evolution_with_caption}) also shows the peak density evolution of the ellipsoidal core as it relaxes to collapse.

A Gaussian sound pulse of half width $\sigma$ contains a wide range of wavelengths, with a characteristic wavelength about twice the full-width, $4\sigma$, and the characteristic frequency $\nu\equiv C_{s,g}/4\sigma$, where $ C_{s,g}$ is the sound speed of the warm gas.  The relative phase can be defined to be proportional to the ratio of the distance of the pulse waist from the core center to the sound traveling distance $L_0(\equiv C_{s,g}/\nu_0)$ in one core oscillation period. Hence, we use these two parameters, $L_0/4\sigma$ and $L/L_0$, to replace the wave frequency and the relative phase of the previous case to search for the threshold pulse amplitudes. To be exact, we define the phase as $\phi\equiv 2\pi(0.34-(L/L_0))$ so that the interaction is most effective at $\phi=0$, and the pulse waist is at $1.276\sigma$ from the pulse peak toward the core. The two numerical values, $0.34$ and $1.276$, are determined after the simulation results are analyzed. In this test, the Gaussian planar sound pulse propagates along the major axis of the ellipsoid.

\begin{figure}
\includegraphics[scale=0.34, angle=270]{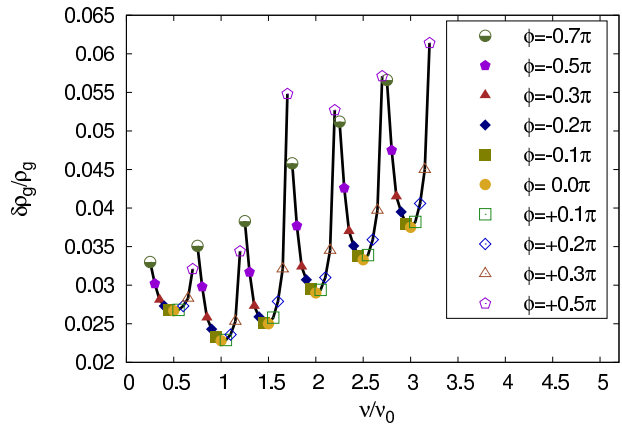}
\caption{Same as Fig. (\ref{fig:phase_diagram_standing_wave}) except for the target being an ellipsoidal core and the driver being a Gaussian pulse. The characteristic frequency of the Gaussian pulse and the arrival phase are defined in the text.}
\label{fig:phase_diagram_Gaussian_wave_irregular_cloud}
\end{figure}

Fig. (\ref{fig:phase_diagram_Gaussian_wave_irregular_cloud}) depicts the result. It follows the general trend as that obtained in the previous case, except for about three times higher threshold wave amplitudes. The ellipsoid creates phase incoherence for the incoming pulse; even when the pulse can reach the core center, the distorted wave front arrives asynchronously, thus requiring a higher wave amplitude to trigger the collapse.  Nevertheless, the dependence on the relative phase between core and wave oscillations is now weaker than the previous case. When the wave frequency meets the broadband resonance condition, the triggered collapse can take place with more than $50\%$ efficiency.

\begin{figure}
\includegraphics[scale=0.355,angle=270]
{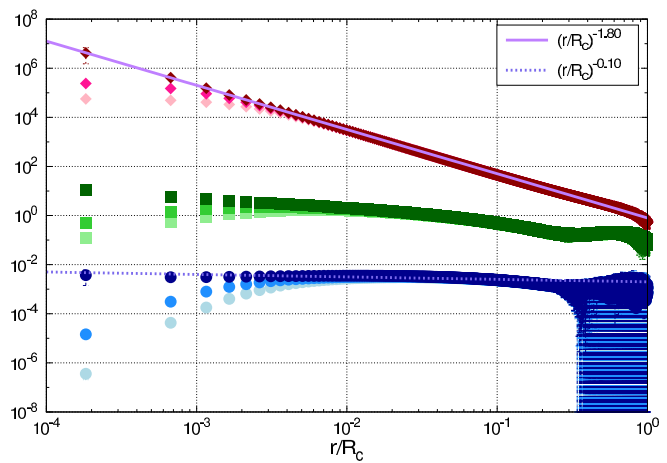}
\caption{A time sequence of shell-averaged radial profiles of density (diamond), infall Mach number (square) and radial mass flux (dot) near the instant of collapse. The time sequence is marked from light symbols to dark symbols. The mass flux is normalized to $100 C_{s,c}^3/G$ for the clarity of display.
The error bars represent deviations from spherical symmetry.}
\label{fig:spherical_average_configuration_during_collapsing_without_caption}
\end{figure}

We also show in Fig. (\ref{fig:spherical_average_configuration_during_collapsing_without_caption}) the radial profiles of density, mass flux and Mach number of the infall for the ellipsoidal core at three instants immediately before the collapse. The density profile fits approximately by a power law with a $-1.80$ logarithmic slope, and the mass flux profile is almost flat with a logarithmic slope $-0.10$. The Mach number exceeds unity at the radius when the mass flux begins to follow the power law, indicating that the power-law mass flux is established in the pressure-free regime. By contrast, the density follows the power-law behavior at a much greater radius when the core pressure is still important. The mass flux, or $\dot{m}$, is about $C_{s,c}^3/G$, comparable to the prediction of the inside-out model \citep{bSh1}, where $C_{s,c}$ is the sound speed at the core boundary. However, the evolution of the core is consistent with the outside-in collapse \citep{bLa, bPe, bHu}. Although profiles for the spontaneously collapsed core described earlier are not shown here but we find that their general behavior is similar to the driven case, but the mass flux is a factor two greater.

\section{Multi-Core Collapse Triggered by a Single Wave}
\label{sec:multi-core collapse triggered by a single wave}

It is interesting to consider the possibility that the passage of a single sound pulse may trigger collapses of multiple cores; this may yield propagating star formation. Since the favorable sound wave has a wavelength few tens greater than the core and since the wave front can extend transversely over a large distance, the wave can in principle interact with a number of cores at different locations and different times. We let a single Gaussian pulse interact with five nearby cores which are placed in a region of size about $1\sigma$. The five cores are identical ellipsoids considered in Sec. (\ref{sec:core collapse triggered by small-amplitude background sound waves}).  The core oscillation frequencies are near the optimal for collapse but the phases are random. 

\begin{figure}
\includegraphics[scale=0.43]{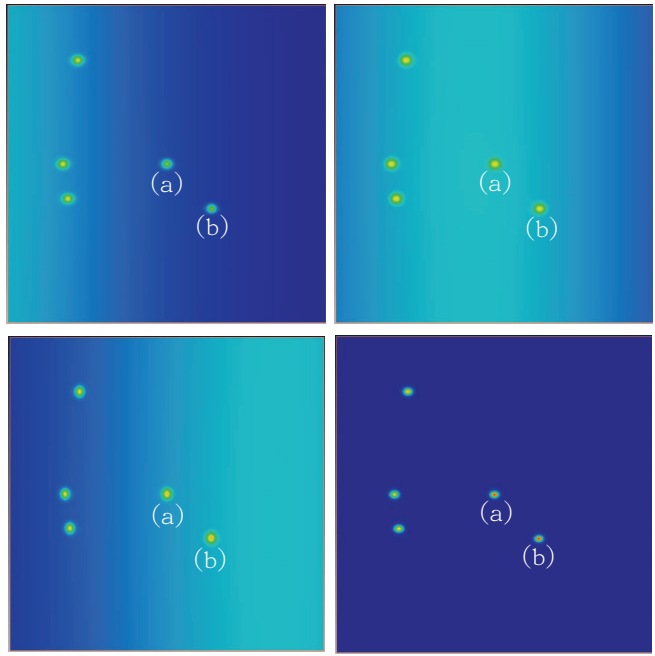}
\caption{Four sequential snapshots of a Gaussian sound wave propagating from left to right interacting multiple ellipsoidal cores. It is noted that two ((a) and (b)), out of five, cores collapse during the wave passage. The five cores have identical oscillation frequencies but differ by the wave encounter phases. }
\label{fig:multi_core_figure}
\end{figure}

The simulation result is shown in Fig. (\ref{fig:multi_core_figure}). As a pulse slightly exceeding the marginal strength arrives, three cores that are out of the favorable phase range do not collapse, but two cores that do collapse.  The two cores collapse in different manners. One (a) that is near the optimal phase collapses promptly, and the other (b) that somewhat deviates from the optimal phase hesitates for a while before it decides to collapse.  Whether these cores collapse or not agrees with the criterion given in Sec. (\ref{sec:core collapse triggered by small-amplitude background sound waves}) for an isolated core as if no other cores were to exist. 

The result may be somewhat puzzling, since back-reactions from the cluster of impinged cores seem not to matter. We find it can be understood as follows. The sound inside the core travels much slower than the Gaussian pulse. Hence any back-reaction wave emanated from the interior of the impinged core will take a finite time to get out of the core, and by the time the perturbation reaches the neighboring cores, the Gaussian wave has already acted on these cores. Hence, the individual core collapse agrees with the criterion given in Sec. (\ref{sec:core collapse triggered by small-amplitude background sound waves}).  

\section{Conclusions}
\label{sec:conclusions}
We consider the linear and nonlinear stabilities of a two-phase gas in the low-mass star forming regions, where a dense core of adiabatic index $\gamma=0.7$ is surrounded by an almost isothermal warm diffuse gas of $\gamma=1.1$ and in between is a blanket of transition layer.

We show that a soft dense core can be stably oscillating as a result of a small average density gradient. A molecular core of $10$ K temperature, 0.2 pc size and 1.5 $M_\odot$ mass yields a typical oscillation period about 2 Myr. However the core will relax and slowly evolves to spontaneous collapse in tens of Myr if left unperturbed. This relaxation of dense core is found possible due to the presence of a vortex flow, which is generated by the very existence of entropy gradient around the soft dense core.

We have also demonstrated a triggered collapse mechanism of a stably oscillating soft core by means of a small-amplitude background wave. This mechanism involves resonant interactions, but the resonance is broad-band, indicative of low-quality factor resonance. The conditions for the triggered collapse are that wave frequencies are comparable to and up to a factor a few higher than the core oscillation frequency, and that the encountering phase should be within about $\pm 90$ degrees of the optimal phase. The latter condition means that this mechanism has about $50\%$ efficiency even when the resonance condition is met.

The triggered collapse mechanism however requires long waves. Sound waves in the warm H\Rmnum{2} region of $10,000$ K must have a wavelength ranging from $6$ to $36$ pc to produce efficient interactions. These small-amplitude long waves may be originated from active sources located tens of pc, or even one hundred pc, away from the star forming regions through the push of ionization fronts and outflows that varies on the time scale of  $0.7-4.5 \times 10^6$ years.

The passage of the sound wave can trigger several collapsing cores at different locations and different times.  Thus, stars may form with tight spatial-temporal correlation tracing the propagation of the wave front, yielding propagating star formation for cores within a range of masses. If the temperature of different cores remains roughly about 10 K, the oscillation period will scale linearly with the core size, and if these cores are not much below the Jeans instability boundary, they will have masses scaled linearly with the core size, thus the oscillation period. A single wave passage can therefore trigger collapse of cores within a mass range slightly short of one order of magnitude.

It follows that young stellar objects (YSO) may have spatial-temporal correlation. This prediction can be tested by measuring the locations, ages and spectral types of YSO in a region of 100 pc. Over this region, the sound transit time is on the order of $10^7$ years. This time scale coincides the life time of YSO before the YSO enters the main sequence, during which the time stamps of various evolutionary stages are distinct, permitting the propagating star formation scenario to be testable.

In this work, we do not consider the effects of rotation and magnetic fields. Our preliminary results show that weak rotation of the core does not affect the triggered collapse mechanism reported in this study; it only changes the central configuration of the collapsed core into two collapsed centers. Magnetic fields may further complicate the central configuration. We will report these effects in a series of future works.

\section*{Acknowledgements}
This project is supported in part by NSC of Taiwan with the grant 100-2112-M-002-018.

\section*{appendix}
\label{sec:appendix}
In Sec. (\ref{sec:stability of a soft core}), we have introduced an optimization parameter $\beta$ to minimize $\delta^2 W$, which yields the optimized $\beta_{op}=G(A,\gamma,\alpha)/2F(A,\gamma,\alpha)$, where $G$ and $F$ would be defined in the next paragraph. We find that $\beta=0$ offers a good approximation to the solution to this problem, and therefore take $\beta=0$ to simplify the presentation in the main text. Here we give a full expression of $\delta^2 W$ for the finite optimized value of $\beta_{op}$.

The full expression of $\delta^2 W$, including $\beta$, can be expressed as
\begin{equation}
\label{equ:stability_with_beta}
\left.\begin{aligned}
\delta^2 W = {{\pi^2G\rho^2_0(1)}\over{R_c}} \Big [ & F(A,\gamma,\alpha)(\beta-\beta_{op})^2 + \\
                                                    & {{4F(A,\gamma,\alpha)H(A,\gamma,\alpha)- G^2(A,\gamma,\alpha)}\over{4F(A,\gamma,\alpha)}} \Big ],
\end{aligned}
\right.
\end{equation}
where $\alpha\equiv T_0(1)/|\Phi_0(1)|$ of Eq. (\ref{equ:stability_condition}),
\begin{equation}
\label{equ:definition_of_F}
\left.\begin{aligned}
F =  & {{96}\over{105}}(3+A)\alpha\gamma+ \gamma \Big ( {{32}\over{55}} + {{83}\over{165}}A + {{31}\over{231}}A^2 \Big ) - \\ 
     & \Big ( {{3904}\over{5775}} + {{53204}\over{86625}}A + {{14956}\over{86625}}A^2 \Big ),
\end{aligned}
\right.
\end{equation}
\begin{equation}
\label{equ:definition_of_G}
\left.\begin{aligned}
G =  & {{18}\over{15}}(3+A)\alpha\gamma+ \gamma \Big ( {{109}\over{140}} + {{433}\over{630}}A + {{241}\over{1260}}A^2 \Big ) - \\
     & \Big ( {{461}\over{525}} + {{1294}\over{1575}}A + {{383}\over{1575}}A^2 \Big ),
\end{aligned}
\right.
\end{equation}
and
\begin{equation}
\label{equ:definition_of_H}
\left.\begin{aligned}
H =  & {{6}\over{15}}(3+A)\alpha\gamma+ \gamma \Big ( {{37}\over{140}} + {{5}\over{21}}A + {{29}\over{420}}A^2 \Big ) - \\
     & \Big ( {{2}\over{7}} + {{139}\over{504}}A + {{31}\over{360}}A^2 \Big ).
\end{aligned}
\right.
\end{equation}

For $\delta^2 W$ to have a minimum with respect to $\beta$, we need $F>0$, and for the marginal stability, we let $4FH-G^2=0$. The condition $4FH-G^2=0$ is found to yield $F>0$ for all positive $A$, $\gamma$ and $\alpha$, hence justifying the $\beta$ parametrization.

\begin{figure}
\includegraphics[scale=0.34, angle=270]{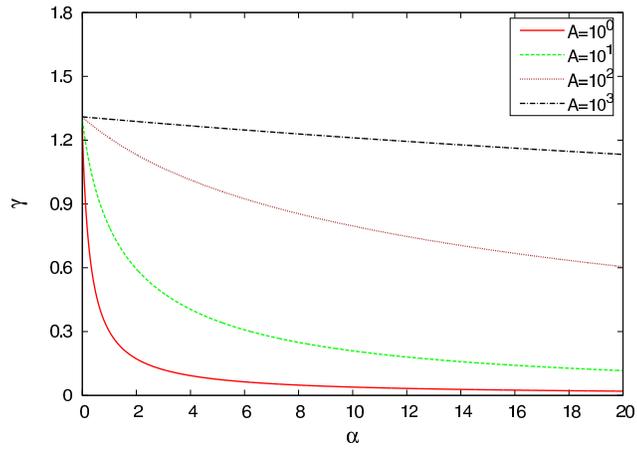}
\caption{Curves of $4FH-G^2=0$ as a function of $\gamma$ and $\alpha$. Different curves represent different values of $A$. The stable region lies above each curve. For $A\to\infty$, $\gamma$ has an upper bound value $1.309$.}
\label{fig:appendix}
\end{figure}

Fig. (\ref{fig:appendix}) plots the family of $4FH-G^2=0$ as a function of $\gamma$ and $\alpha$ with $A$ as the family parameter. Above each curve, $4FH-G^2>0$, representing the stable region. As one can see, the family of curves asymtote to a straight line for which $A\to\infty$ and $\gamma$ has a limiting value $1.309$.

This stability criterion indicates that when $\gamma$ increases, harder cores become more stable and one needs a larger $A$ for the core to be unstable. However the stability criterion can become invalid when $\gamma = 1.309$, where it predicts that the marginal stable $A$ is infinity, and when $\gamma$ is greater than this value, the core becomes completely stable. Clearly the result cannot be correct. The number $\gamma=1.309$ should not be taken literally, but the tendency of a hard core for the above stability analysis to be invalid may be understood as follows. For a hard core, the equilibrium density $\rho(r)$ tends to cross zero at a finite radius, the profile is convex and hence it should be better approximated by a quadratic model, e.g., $\rho_0(x)\sim 1+(A-1)(1-x^2)$. On the other hand for a soft core, the equilibrium density tends to extend to infinity, the profile is concave in most part of the mass and may be captured by the linear model. For example only when $\gamma \leq 1.2$, the Lane-Emden equation permits a power-law density $\rho(r) \sim r^{[-2/(2-\gamma)]}$ \citep*{bBi}, the profile is concave and we think that Eq.(\ref{equ:stability_condition}) can be approximately valid. Extension of the above approximate treatment of linear stability analysis to a hard core is a interesting problem in other applications and worths further investigations.

\label{lastpage}


\begin{thebibliography}{99}

\bibitem[\protect\citeauthoryear{Binney \& Tremaine}{2008}]{bBi} Binney J., Tremaine S., 2008, Galactic Dynamics (2nd ed.). Princeton Univ. Press, Princeton and Oxford, p. 302
\bibitem[\protect\citeauthoryear{Bachiller, Guilloteau \& Kahane}{Bachiller et al.}{1987}]{bBa} Bachiller R., Guilloteau S., Kahane C., 1987, A\&A, 173, 324
\bibitem[\protect\citeauthoryear{Brenner, Hilgenfeldt \& Lohse}{Brenner et al.}{2002}]{bBr} Brenner M. P., Hilgenfeldt S., Lohse D., 2002, Rev. Mod. Phys., 74, 425
\bibitem[\protect\citeauthoryear{Hunter}{1977}]{bHu} Hunter C. 1977, ApJ, 218, 834
\bibitem[\protect\citeauthoryear{Klein, McKee \& Colella}{Klein et al.}{1994}]{bKl} Klein R. I., McKee C. F., Colella P., 1994, ApJ, 420, 213
\bibitem[\protect\citeauthoryear{Lada et al.}{2003}]{bLada} Lada C. J., Bergin E. A., Alves J. F, Huard T. L., 2003, ApJ, 586, 286
\bibitem[\protect\citeauthoryear{Larson}{1969}]{bLa} Larson R. B., 1969, MNRAS, 145, 271
\bibitem[\protect\citeauthoryear{Li \& Nakamura}{2006}]{bLi} Li Z. Y., Nakamura F. 2006, ApJ, 640, L187
\bibitem[\protect\citeauthoryear{Mac Low \& Klessen}{2004}]{bMa} Mac Low M. M., Klessen R. S., 2004, Rev. Mod. Phys., 76, 125
\bibitem[\protect\citeauthoryear{Nakamura \& Li}{2007}]{bNa1} Nakamura F., Li Z. Y., 2007, ApJ, 662, 395
\bibitem[\protect\citeauthoryear{Nakamura et al.}{2006}]{bNa2} Nakamura F., McKee C. F., Klein R. I., Fisher R. T., 2006, ApJS, 164, 477
\bibitem[\protect\citeauthoryear{Penston}{1969}]{bPe} Penston M. V., 1969, MNRAS, 144, 425
\bibitem[\protect\citeauthoryear{Redman, Keto \& Rawlings}{Redman et al.}{2006}]{bRe} Redman M. P., Keto E., Rawlings J. M. C., 2006, MNRAS, 370, L1
\bibitem[\protect\citeauthoryear{Shu}{1977}]{bSh1} Shu F. H., 1977, ApJ, 214, 488
\bibitem[\protect\citeauthoryear{Spaans \& Silk}{2000}]{bSp} Spaans M., Silk J., 2000, ApJ, 538, 115
\bibitem[\protect\citeauthoryear{Stone \& Norman}{1992}]{bSt} Stone J. M., Norman M. L., 1992, ApJ, 390, L17
\bibitem[\protect\citeauthoryear{Schive, Tsai \& Chiueh}{Schive et al.}{2010}]{bSc1} Schive H. Y., Tsai Y. C., Chiueh T. H., 2010, ApJS, 186, 457
\bibitem[\protect\citeauthoryear{Schive, Zhang \& Chiueh}{Schive et al.}{2012}]{bSc2} Schive H. Y., Zhang, U. H., Chiueh T. H., 2012, IJHPCA, 26, 367
\bibitem[\protect\citeauthoryear{V\'{a}zquez-Semadeni, Gazol \& Scalo}{V\'{a}zquez-Semadeni et al.}{2000}]{bVa} V\'{a}zquez-Semadeni E., Gazol A., Scalo J. 2000, ApJ, 540, 271
\bibitem[\protect\citeauthoryear{Xu \& Stone}{1995}]{bXu} Xu J., Stone J. M., 1995, ApJ, 454, 172

\end{thebibliography}
\end{document}